\documentclass[a4paper]{article}
\usepackage[english]{babel}
\usepackage[utf8]{inputenc}
\usepackage[T1]{fontenc}
\usepackage[usenames, dvipsnames]{xcolor}
\usepackage{tikz}
\usepackage{amsthm}
\usepackage{amsmath} 
\usepackage{fullpage}
\usepackage{hyperref}
\usepackage{xspace,float,caption}
\usepackage{a4wide}
\usepackage[utopia]{mathdesign} 
\usepackage[color=green!30]{todonotes}
\usepackage{authblk}

\everymath{\displaystyle}
\hypersetup{
	colorlinks=true,
	linkcolor=blue,
	filecolor=blue,      
	urlcolor=blue,
	citecolor=blue
}

\newtheorem{theorem}{Theorem}

\newtheorem{lemma}[theorem]{Lemma}
\newtheorem{corollary}[theorem]{Corollary}
\newtheorem{definition}[theorem]{Definition}
\newtheorem{property}[theorem]{Proposition}

\newtheorem{claim}{Claim}[theorem]

\DeclareMathOperator{\nbIS}{\#IS}
\DeclareMathOperator{\nbWIS}{\#WIS}
\DeclareMathOperator{\cwd}{cwd}
\DeclareMathOperator{\poly}{poly}

\newcommand{\tplus}{\mathbin{\tilde +}}
\newcommand{\ttimes}{\mathbin{\tilde \times}}
\newcommand{\tdiv}{\mathbin{\tilde /}}

\newcommand{\indepSet}{\mathcal{I}}
\newcommand{\vecOne}{\overline{\textbf{1}}}

\newcommand{\an}[1]{\langle#1\rangle}
\newcommand{\cI}{\mathcal{I}} 
\newcommand{\es}{\varnothing}
\newcommand{\ie}{\textsl{i.e.\xspace}} 
\newcommand{\IQ}{\mathbb{Q}} 
\newcommand{\LE}{$\Lambda$-expression}
\newcommand{\NP}{\textsf{NP}}
\newcommand{\numP}{\textsf{\#P}}
\newcommand{\RP}{\textsf{RP}}
\newcommand{\PT}{\textsf{P}} 
\newcommand{\sm}{\setminus}

\title{Counting independent sets in strongly orderable graphs}
\author[1]{Marc Heinrich\footnote{Work supported by EPSRC grants EP/S016562/1, “Sampling in hereditary classes”.}}
\author[1]{Haiko M\"uller*}
\affil[1]{School of Computing, University of Leeds, Leeds LS2 9JT, UK}
\date\today

\begin{document}
	\maketitle
	\begin{abstract}
		We consider the problem of devising algorithms to count exactly the number of independent sets of a graph $G$. We show that there is a polynomial time algorithm for this problem when $G$ is restricted to the class of strongly orderable graphs, a superclass of chordal bipartite graphs. We also show that such an algorithm exists for graphs of bounded clique-width. Our results extends to a more general setting of counting independent sets in a weighted graph and can be used to count the number of independent sets of any given size $k$.
	\end{abstract}

	\section{Introduction}
	
	Sometimes, knowing that there is a solution to a particular problem is not enough, and we would like instead to compute how many solutions there are. Ever since the work of Valiant~\cite{Val79} introducing the complexity class \numP\ which captures this type of problems, the counting version of many problems have been studied such as for example constraint satisfaction problems~\cite{Bul08}, spanning trees~\cite{BP83}, independent sets~\cite{Vad01}, colourings~\cite{BDGJ99},\ldots

	Counting problems have been motivated in part by reliability questions. Indeed, the resilience of a network to certain attacks is connected to the number of subgraphs which satisfy certain properties~\cite{BP83}, such as being connected for example.
	These problems are also motivated by some questions from statistical physics where models of particle interactions lead to computing some quantities called partition functions, which are often weighted generalisations of some form of counting problem. Finally, counting is also closely tied to the problem of random sampling~\cite{JVV86}: drawing a solution to a problem according to a certain (often uniform) distribution.
	
	In this paper, we are interested in the problem of counting the number of independent sets in a graph. Ever since the work of~\cite{PB83}, we know that this problem is \numP-hard in general graphs. Many other results, both positive and negative were obtained on this problem for more restricted classes of graphs~\cite{Gre00, OUU08, FG04, DM19}, and this paper follows into this direction by studying the complexity of the problem on two different classes of graphs. Our main result is a an algorithm to compute exactly the number of independent sets in strongly orderable graphs, a superclass of chordal bipartite. This class was introduced by Dagran~\cite{Dag00} as a way to study how the existence of particular elimination ordering in a graph influenced the complexity of various problems. Our second result is a dynamic programming algorithms to compute the number of independent sets in graphs of bounded clique-width. This class includes graphs of bounded treewidth, as well as other non-sparse classes of graphs such as cographs and distance hereditary graphs.

	\paragraph{Existing work.}
	There are several significant results on the problem of counting independent sets in a graph. The problem was first shown to be \numP-hard, even for bipartite graphs in the work of Provan and Ball~\cite{PB83}. This result was later improved in several papers showing that the hardness holds for other classes of graphs such as graphs of maximum degree~3~\cite{Gre00}, planar graphs~\cite{Vad01} or comparability graphs~\cite{DM19}. From the point of view of parametrized complexity, counting the number of independent sets of a given size was shown to be $W[1]$-hard~\cite{FG04}, and hence is unlikely to have an FPT algorithm. 
	
\begin{figure}
	\begin{center}
          \tikzset{ns/.style={draw, rectangle, rounded corners = 3pt, align=center}}
	\begin{tikzpicture}[xscale=2.5, yscale=1.2]
	\node[ns] (wCh) at (3, 0.0) {weakly chordal} ;
	\node[ns] (bip) at (5, 2.0) {bipartite~\cite{PB83}} ;
	\node[ns] (sOr) at (3,-1.5) {strongly \\ orderable \\ \textbf{[\hyperref[sec:strongOrd]{here}]} } ;
	\node[ns] (cbg) at (3,-2.75) {chordal bipartite} ; 
	\node[ns] (tCo) at (5, 1.0) {tree-convex \\ bipartite \cite{LC17b}} ;
	\node[ns] (bPe) at (3,-4.0) {bipartite permutation \\ \cite{LC17}} ; 
	\node[ns] (pla) at (1, 1.5) {planar \\ \cite{Vad01}} ;
	\node[ns] (coc) at (5,-1.5) {cocomparability \\ \cite{DM19}} ; 
	\node[ns] (trz) at (5,-4.0) {trapezoid \\ \cite{LC09}} ; 
	\node[ns] (bcw) at (1,-2.5) {bounded $\cwd$ \\ \textbf{[\hyperref[sec:boundedCW]{here}]}} ; 
	\node[ns] (dhg) at (1,-4.0) {distance \\ hereditary \\ \cite{Lin18}} ; 
	\node[ns] (cho) at (1,-1.3) {chordal \\ \cite{OUU08}} ;
	
        \draw (bPe) -- (cbg) -- (sOr) -- (wCh);
        \draw (bPe) -- (coc) -- (trz) -- (wCh);
	\draw (bcw) -- (dhg) -- (wCh) -- (cho);
        \draw (cbg) -- (tCo) -- (bip);
	
	\draw[line width = 3pt, black, draw opacity =0.5] (0.3, 0.5) -- (6, 0.5) ;
	\node[anchor=south west] at (0.3, 0.5) {\scriptsize \textsf{PSPACE}-complete} ;
	
	\draw[line width = 3pt, black, draw opacity =0.5] (0.3, -0.5) -- (6, -0.5) ;
	\node[anchor=north west] at (0.3, -0.5) {\scriptsize Polynomial} ;
	\end{tikzpicture}
	\end{center}
	
	\caption{\label{fig:results}Complexity landscape for the problem of counting exactly the number of independent sets in graphs. The complexity for weakly chordal graphs is still open.}
	\end{figure}
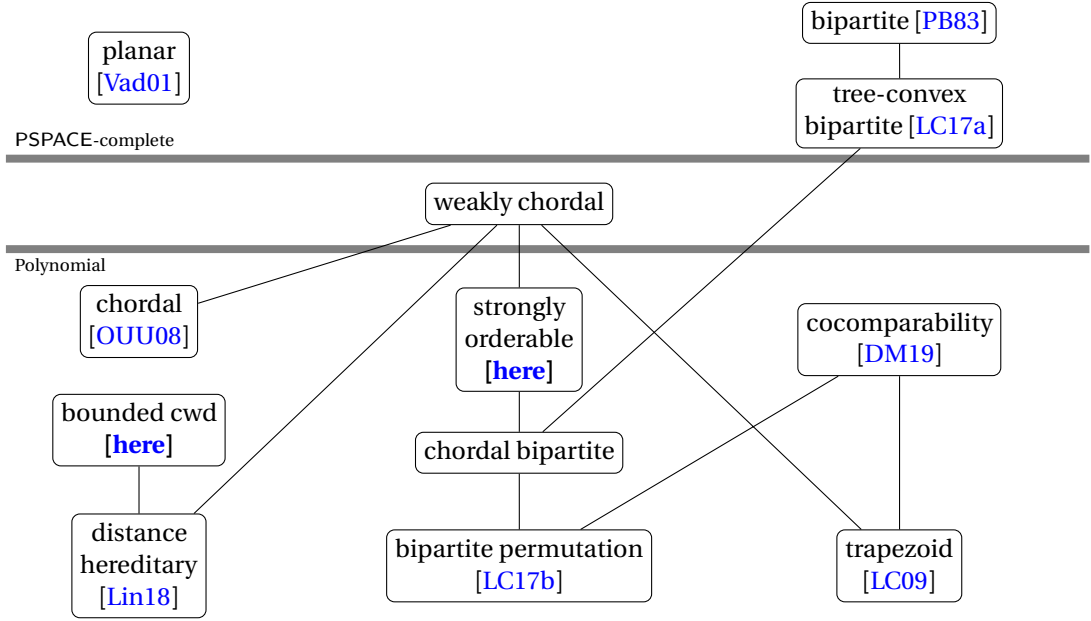

	On the positive side, the problem has a polynomial time algorithm for several more restricted classes of graphs such as chordal graphs~\cite{OUU08}, cocomparability graphs~\cite{DM19}, graphs with bounded tree-width~\cite{WTZL18} or tolerance graphs~\cite{LS15}, just to give a few examples. A picture representing all the known results, and the relations between the different classes can be found in Figure~\ref{fig:results}.
	
	The problem of counting the number of matchings in a graph, \ie, independent sets in the linegraph, has also been widely studied. Counting the number of perfect matchings in bipartite graphs was one of the first problem shown to be \numP-complete by Valiant in her paper~\cite{Val79} which introduced \numP\ as a complexity class. This result was later extended to counting all matchings (not just the perfect ones)~\cite{Val79b}. Similarly to independent sets, the problem was studied for several classes of graphs, and was shown to be \numP-hard on chordal and chordal bipartite graphs~\cite{OUU09}. Surprisingly, while it is also hard on planar graphs~\cite{DM19}, the problem of counting only perfect matchings for this class is related to Pfaffian orientations and admits a polynomial time algorithm~\cite{Kas63}.

	The problem of counting the number of independent sets in a graph has also been studied from the point of view of approximation. Most of the results on the problem come in fact from statistical physics and the study of a particular model of particle interactions called the hard core model. This model can be seen as a weighted generalisation of counting the independent sets in the graph. More precisely, given a parameter $\lambda$, each independent set $X$ has an associated weight of $\lambda^{|X|}$, and the partition function for the hard core model with fugacity $\lambda$ corresponds to the following quantity:
	\[ \sum_{S \in \indepSet(G)} \lambda^{|S|}\;, \]
	where $\indepSet(G)$ is the set of all the independent sets of the graph $G$. We can observe that when the parameter $\lambda$ is equal to $1$, this is exactly counting the number of independent sets of $G$. We can also remark that the quantity above can be seen as a polynomial in $\lambda$, and the coefficients of this polynomial are the numbers of independent sets of $G$ of a given size $k$ for all possible values of $k$. Hence, asking to compute exactly this quantity for any parameter $\lambda$, is essentially the same as computing the number of independent sets of size $k$ of the graph, for any possible $k$.
	
	From the point of view of approximation, the complexity of this problem on graphs of maximum degree $\Delta$ is well understood. Indeed, there is a parameter $\lambda_{\mathrm c}(\Delta) := (\Delta-1)^{\Delta-1} / (\Delta-2)^\Delta$, which has some physical interpretation (see~\cite{Wei06, GSV16} for example), and the complexity of the problem depends only on whether $\lambda < \lambda_{\mathrm c}$ or not. When $\lambda < \lambda_{\mathrm c}$, the problem admits a FPTAS (Fully Polynomial-Time Approximation Scheme)~\cite{Wei06}, \ie, an algorithm which computes a $(1+\varepsilon)$-approximation of the result in time polynomial in both the size of the instance and $1/\varepsilon$. On the other hand, when $\lambda > \lambda_{\mathrm c}$, no such approximation scheme exists, even if randomization is allowed, and assuming $\NP \neq \RP$ (this is the equivalent of $\PT$ versus $\NP$ for randomized algorithms)~\cite{GSV16, Sly10, MWW09, GGSVY11}. Remark that when $\lambda = \lambda_{\mathrm c}$ is exactly at the threshold the complexity is apparently still open. Note moreover that $\lambda_{\mathrm c}$ is larger than $1$ when $\Delta \leq 5$, and is smaller than $1$ when $\Delta \geq 6$. This implies that the number of independent sets of a graph of maximum degree $\Delta$ can be approximated in polynomial time when $\Delta$ is at most $5$, but does not admit an approximation scheme when $\Delta$ is $6$ or more. Finally, a last interesting fact is that approximating the number of independent sets for the class of bipartite graphs appears to have an intermediate complexity. More precisely, it is believed~\cite{CGG+16} that it does not admit an approximation scheme, nor is it AP-reducible to \textsf{\#SAT}, the problem of counting the number of satisfiable assignment of a boolean formula. This claim is supported by the fact that many problems were shown to be equivalent to approximating the number of independent sets in a bipartite graph, and they form a class sometimes called \textsf{\#BIS}.

	\paragraph{Our result.}
	The main result of this paper is a proof that the number of independent sets of a graph can be computed in polynomial time for strongly orderable graphs. This class (that we define in the next section) is a subclass of weakly chordal graphs, and contains the chordal bipartite graphs. In fact, our algorithm applies to a slightly more general problem. Given a weight function $w$ on the vertices of the graph $G$, our goal is to compute the following quantity:
	
	\[ \nbWIS(G, w) = \sum_{S \in \indepSet(G)} \prod_{v \in S} w(v)\;. \]
	
	Note that the number of independent sets of a graph corresponds to the case where all the weights are equal to $1$. The partition function for the hard core model with fugacity $\lambda$ corresponds to the case where all the weights are equal to $\lambda$. This quantity can be interpreted as a more general version of the partition function of the hard core model where each vertex $v$ is given an individual fugacity of $w(v)$. Our main result is the following theorem:
	
	\begin{theorem}
		\label{thm:strongOrd}
		$\nbWIS(G, w)$ can be computed in polynomial time if $G$ is a strongly orderable graph, and $w$ is an arbitrary positive weight function.
	\end{theorem}
	
	By the remark above, the case of counting the number of independent sets, or computing the partition function for this class of graph is also polynomial since these are special cases of the result above. If we compare this result to the problem of counting matchings, it appears to be somewhat surprising. Indeed, it is known that counting independent sets in bipartite graphs is \numP-hard, and even the approximation of the problem seem unlikely to have a polynomial time approximation scheme. On the other hand, there are polynomial time approximation schemes for matchings in bipartite graphs, and counting the number of matching exactly is \numP-hard. Hence, even though the counting problem appears to be easier for matchings than for independent sets on bipartite graphs, it is in fact harder when we restrict further to the class of chordal bipartite graphs. Our result proves that the situation for strongly orderable graphs is similar to that of chordal graphs in the sense that counting independent sets on this class is polynomial, while counting matchings is \numP-hard.
	
	In addition, our result is a significant improvement on~\cite{LC17} which proved that a polynomial-time algorithm existed for computing the number of independent sets for bipartite permutation graphs, a subclass of chordal bipartite graphs. We also show that problem is polynomial for graphs of bounded clique-width:
	\begin{theorem}
		\label{thm:boundedCW}
		$\nbWIS(G, w)$ can be computed in polynomial time if $G$ has bounded bounded clique-width, and $w$ is an arbitrary positive weight function.
	\end{theorem}
	This generalizes at the same time a result from~\cite{WTZL18} which proved it for graphs of bounded tree-width, and a result from~\cite{Lin18}, which concerned distance-hereditary graphs which have clique-width at most $3$. Again our algorithm works for the more general problem of the hard core model with fugacity $\lambda$, even when the fugacity is different for each vertex.
	
	\paragraph{Overview} The rest of the paper is organized as follows. In Section~\ref{sec:defs} we give some basic definitions from graph theory and formally define our problem. In Section~\ref{sec:tech} we first give an overview on how the proof for strongly orderable graphs works, and we then prove two technical lemmas which are used later in the proof. Section~\ref{sec:strongOrd} contains the description of the algorithm and the proof for strongly orderable graphs. Finally, Section~\ref{sec:boundedCW} proves our result for graphs of bounded clique-width.
	
	\section{Problem definition and notations}
	\label{sec:defs}
	
	\subsection{Graphs and graph classes}
	
	We start with some basic definitions and notations from graph theory, as well as the definitions of some classes of graphs that will be used in the rest of the paper.
	
        For subsets $U,W \subseteq V$, let $U \udtimes W := \{\{u,w\} \mid u \in U \text{~and~} w \in W\sm\{u\}\}$ denote the set of unordered pairs with one element in $U$ and one in $W$. A graph $G=(V,E)$ is described by a finite set of vertices $V$, and a set of edges $E \subseteq V \udtimes V$. Given two vertices $u,w \in V$, we will denote by $uw := \{u,w\}$ the edge between $u$ and $w$. Unless stated otherwise all the graphs in this paper are simple, there are no loops and no multiple edges, and undirected; if $uw \in E$ then $wu \in E$.
	
	Given a graph $G= (V,E)$, and a set of vertices $U \subseteq V$, we denote by $G[U]$ the subgraph induced by the vertices in $U$, \ie, $G[U] := (U, E \cap (U \udtimes U))$. The notation $G \sm U$ is shorthand for $G[V \sm U]$ and for $v \in V$ we abbreviate $G \sm \{v\}$ by $G \sm v$. If $U$ and $W$ are disjoint subset of vertices, then $G[U,W]$ denotes the bipartite graph with colour classes $U$ and $W$, and edges $E \cap (U \udtimes W)$. Given a vertex $v$, we denote by $N_G(v)$ the neighbours of $v$, and $N^2_G(v)$ the vertices at distance $2$, and we will drop the subscript $G$ when the graph is clear from the context.

        A \textbf{weight function} $w$ on $G$ is an assignment of non-negative rational weights to the vertices of $G$. We can see it as a function $w : V \to \IQ$, but also as a vector of size $|V|$. We denote by $\vecOne$ the weight functions equal to $1$ for all the vertices of the graph. If $H$ is a subgraph of $G$, then we will denote by $w|_H$ the restriction of $w$ to the vertices of $H$.
	
	The first class of graph that we consider are the chordal bipartite graphs, which are the bipartite equivalent of chordal graphs defined as follows:
	
	\begin{definition}
		A graph $G$ is \textbf{chordal bipartite} if it is bipartite and does not contain any induced cycle of length $5$ or more.
	\end{definition}
	
	Similarly to chordal graphs, there are other characterizations of chordal bipartite graphs based on some particular ordering of its vertices, however these properties will not be needed in the rest of the paper. Instead we focus on a special subclass of chordal bipartite called chain graphs with the following definition:
	\begin{definition}[\cite{Yan82}]
		A bipartite graph $G$ is a \textbf{chain graph} if for every vertices $u$ and $v$ on the same side of the bipartition, we have either $N(u) \subseteq N(v)$ or the contrary $N(v) \subseteq N(u)$.
	\end{definition}
	Chain graphs graphs will play an important role in the proof of our main result which concerns graphs called strongly orderable. The latter class is a generalisation of both strongly chordal graphs and chordal bipartite graphs. 
	
	\begin{definition}[\cite{VBW92}]
		A graph $G$ is \textbf{strongly orderable} if there exists an ordering of the vertices $v_1, \ldots, v_n$ such that for all $i <j $ and $k <l$, if $v_iv_k, v_iv_l$ and $v_kv_j$ are edges of $G$, then $v_jv_l$ is also an edge of $G$. An ordering satisfying this property is called a \textbf{strong ordering}.
	\end{definition}
	Moreover, graphs in this class can be recognized in polynomial time, and a strong ordering can be computed in time $O(|E|\cdot|V|)$~\cite{Dra00}.
	Finally, the class of cographs will be used in the proofs.
	\begin{definition}
		\label{def:cographs}
		\textbf{Cographs} are graphs constructed starting from single vertices by the operations of:
		\begin{itemize}
			\item disjoint union: if $(V, E)$ and $(W, F)$ are cographs, then $(V \cup W, E \cup F)$ is a cograph;
			\item complete join: if $(V, E)$ and $(W, F)$ are cographs, then $(V \cup W, E \cup F \cup (V \udtimes W))$ is a cograph.
		\end{itemize}
	\end{definition}
	There are several other ways to characterize cographs. We will only need the following property in the rest of our proof:
	\begin{property}[\cite{CLS81}]
		\label{prop:cographs}
		A graph is a cograph if and only if it does not contain $P_4$, a path on four vertices, as an induced subgraph.
	\end{property}
        For further information on graphs in these restricted classes, their properties and references to original work see \cite{Gol80, BLS99}.
	We will also need the concept of clique-width. The formal definition of this parameter can be found in Section~\ref{sec:boundedCW}.

	\subsection{Arithmetic circuit}
	Throughout this paper, an \textbf{arithmetic circuit} over the set of variables $x_1, \ldots, x_n$ is a directed acyclic graph such that each vertex is either:
	\begin{itemize}
		\item A vertex of in-degree zero, and labelled with either $x_i$ for some $i \geq 0$, or some fixed constant $z \in \IQ$, 
		\item A vertex of in-degree two, and labelled with one of the four following operations: $+, -, \times, /$.
	\end{itemize}
	Note that in the case of subtraction and division gates, the order of the two inputs is important. Remark that this is not the standard definition of arithmetic circuit, as they usually do not include division gates. We include them here because they will play an important role in our proofs.
	A circuit $C$ describes a function $\IQ^n \rightarrow \IQ^k$, where $k$ is the number of outputs (\ie, vertices with out-degree zero) which is a rational fraction: it is the quotient of two polynomial functions. In the following, a circuit will be called \textbf{positive} if it does not contain any subtraction gates, and all the constants in the circuit are positive. Positive circuits will play an important role as they behave nicely from the point of view of approximate computations.
	
	\subsection{Problem definition}
	
	Given a graph $G$, we denote by $\nbIS(G)$ the number of independent sets in $G$. We will consider a more general problem of computing a weighted number of independent sets. Given a graph $G$, and a weight functions $w$, we will denote by $\nbWIS(G, w)$ the number of weighted independent sets defined by the following quantity:
	
	\[ \nbWIS(G, w) = \sum_{S \in \indepSet(G)} \prod_{v \in S} w(v) \;. \]
	
	For a fixed graph $G$, the function $w \mapsto \nbWIS(G, w)$ is a polynomial on $n$ variables whose maximum degree is the size of the largest independent set in $G$. We will denote by $P_G$ this polynomial, which is sometimes called the independence polynomial of the graph $G$~\cite{GH83}. Our goal is to investigate for which classes of graphs this quantity can be computed exactly. It is easy to see that computing the number of independent sets corresponds to the case where the weight function $w$ is $\vecOne$. More generally, computing the partition function of the hard core model with fugacity $\lambda$ is also a special case of this problem where the weight function $w$ is $\lambda \cdot \vecOne$.
	
	
	\section{Outline and technical results}
	\label{sec:tech}
		
	The proof of Theorem~\ref{thm:strongOrd} proceeds in essentially two steps. The first step consists in building an arithmetic circuit computing the independence polynomial $P_G$ by using the structural properties of strongly orderable graphs described in Lemma~\ref{lem:strongOrd} below. The input to this circuit are the different coefficients of the weight function $w$, and it outputs $\nbWIS(G,w)$ for the input graph $G$. Once this arithmetic circuit is built, one can think that evaluating it for a specific weight function $w$ would be straightforward, unfortunately some technicalities force us to take a slight detour. Indeed, a direct evaluation of the circuit might not be possible since intermediate results might required a very large (super-polynomial) number of bits to be represented exactly. On the other hand, the presence of division gates implies that the standard trick of making all the computation modulo some (large enough) prime $p$ does not work since this might lead to divisions by zero if we were not careful in the choice of the prime $p$.
	
	Instead, our proof proceeds by first observing that the circuit built in the previous step is positive, and then use this property to evaluate the circuit approximately, up to a multiplicative  $(1+\varepsilon)$ factor. Finally, since the polynomial $P_G$ computed by this circuit has a degree at most $n$, the exact value can be recovered from the approximation by a simple rounding, provided $\varepsilon$ is chosen sufficiently small (but still requiring only a polynomial number of bits to represent).

	We now proceed by proving two technical lemmas used in the proof of our theorem. The first one, Lemma~\ref{lem:strongOrd} below gives the structural properties of strongly orderable graphs that we will need in the first step of the proof. The second one concerns the evaluation of positive arithmetic circuits using approximations which is used in the second step of the proof.
	
	\begin{lemma}
		\label{lem:strongOrd}
		Let $G$ be a strongly orderable graphs, and $u$ be the first vertex in a strong ordering of $G$. Then:
		\begin{itemize}
			\item $G[N(u)]$ is a cograph,
			\item $G[N(u), N^2(u)]$ is a chain graph.
		\end{itemize}
	\end{lemma}
	\begin{proof}
		Let us first consider the first point. Let us assume by contradiction that $N(u)$ is not a cograph.  By Proposition~\ref{prop:cographs} this means $N(u)$ contains an induced path $P$ on four vertices. Let $v_1, \ldots, v_4$ be the vertices in this path. Up to symmetry, we can assume that either $v_1$ or $v_2$ is the first vertex of $P$ according to the strong ordering of~$G$. We consider these two cases:
		\begin{itemize}
			\item $v_1$ is the first vertex of $P$ according to the strong ordering of $G$. In this case, we know that $uv_1$, $uv_4$ and $v_1v_2$ are edges of $G$, and $u$ appears before $v_2$, and $v_1$ appears before $v_4$ in the strong ordering of $G$. Hence it follows that $v_2v_4$ is an edge of $G$ by the property of the strong ordering, a contradiction of the assumption that $P$ is an induced path.
			\item $v_2$ is the first vertex of $P$ according to the strong ordering of $G$. In this case, we know that $uv_2$, $uv_4$ and $v_2v_1$ are edges of $G$, and $u$ appears before $v_1$, and $v_2$ appears before $v_4$ in the strong ordering of $G$. Hence it follows that $v_1v_4$ is an edge of $G$, contradicting again the assumption that $P$ is an induced path.
		\end{itemize}
		
		In both cases we obtain a contradiction, hence the first property holds.
		Let us now prove the second point. Let $u_1, u_2$ be two vertices in $N(u)$ such that $u_1$ appears before $u_2$ in the strong ordering of~$G$. Let $v_1$ be a neighbour of $u_1$ in $H := G[N(u), N^2(u)]$. Then by construction, we know that $uu_1$, $uu_2$ and $u_1v_1$ are edges of $G$, and $x$ appears before $v_1$, and $u_1$ appears before $u_2$ in the strong ordering of $G$. By the property of the strong ordering, it follows that $u_2v_1$ is an edge of $G$. Since this holds for every neighbour of $u_1$ in $H$, it follows that $N_H(u_1) \subseteq N_H(u_2)$. Since this holds for any pair of vertices $u_1, u_2$ in $N(u)$, it implies that $H$ is a chain graph.
	\end{proof}
	
	The second technical result below allows us to evaluate positive arithmetic circuits. 
	
	\begin{lemma}
		\label{lem:computeCircuit}
		Let $C$ be a positive arithmetic circuit with $n$ inputs and one output computing a polynomial $P$ of degree at most $d$. Let $x$ a positive rational vector of dimension $n$ such that both $x$ and $C(x)$ can be represented using $n_{\mathrm b}$ bits. Then $C(x)$ can be computed exactly in time polynomial in $|C|, n, d$ and $n_{\mathrm b}$.
	\end{lemma}
	Note that in the lemma above, the fact that the circuit is positive, and that $x > 0$ is important. Indeed, without these conditions, when doing the computations using floating point arithmetic, an important loss of relative precision can occur when subtracting two quantities of close values. Remark also that a direct evaluation of the circuit is not necessarily possible. Indeed, because there are some division gates, it is possible for the intermediate results to require more than a polynomial number of bits to be represented exactly, even if the end-result does not.
	
	\begin{proof}
		We will use floating point arithmetic to evaluate the circuit $C$ on the input $x$. In the rest of the proof, $b_{\mathrm m}$ will represent the number of bits we use for the mantissa of the floating point numbers, and $b_{\mathrm e}$ is the number of bits used for the exponent. Since we know that each coefficient of $x$ can be written using only $n_{\mathrm b}$ bits, it follows that the coefficient of $x$ satisfy that for all $i \leq n$: $2^{-n_{\mathrm b}} \leq x_i \leq 2^{n_{\mathrm b}}$. Moreover, we make the following claim:
		\begin{claim}
			All the intermediate values resulting from computing $C(x)$ are between $\frac 1 A$ and $A$ where $A = 2^{n_{\mathrm b} \cdot 2^{|C|}}$.
		\end{claim}
		\begin{proof}
			We prove the result by induction on $|C|$. If $|C| = 0$, then the result holds immediately. Otherwise, consider the circuit $C'$ obtained from $C$ by removing one of the output gates (\ie, with out-degree zero). Using the induction hypothesis, all the intermediate values of $C'$ have size at most $2^{n_{\mathrm b} 2^{|C|-1}}$. If the gate we just removed is an addition gate, then its output is at most $2^{n_{\mathrm b} 2^{|C|-1} +1}$. If it is a multiplication or a division gate, then it is at most $(2^{n_{\mathrm b} 2^{|C|-1}} )^2$. If it is a gate with in-degree zero, then the result holds immediately.
			The lower bounds holds using a similar argument.
		\end{proof}
		
		A consequence of this fact is that if we use at least $b_{\mathrm e} \geq \log(n_{\mathrm b}) + |C| + 1 > \log(A)$ bits for the exponent, all the computation can be done using floating point arithmetic without any risk of overflow. Hence, the only thing that remains to be done is to investigate how large the mantissa needs to be to get a sufficient precision and retrieve the exact result in the end. 
		
		Let $\varepsilon =  2^{-b_{\mathrm m}}$, and let $K = \frac {1 + \varepsilon} {1-\varepsilon}$. Let us denote by $\tilde +, \tilde \times, \tilde /$ the operations $+, \times, /$ on floating point numbers. We know that for any operator $\circ \in \{+, \times, /\}$, and any values $a$ and $b$ represented as floating point numbers, we have $(1-\varepsilon) (a \circ b) \leq a \mathbin{\tilde{\circ}} b \leq (1+\varepsilon)(a \circ b)$. In other words, this means that every time we perform a floating point operation, we loose at most a $1\pm\varepsilon$ factor in the relative precision.

		 We will show by induction on $|C|$ that for all intermediate value $y$ in the computation of $C(x)$, if $\tilde{y}$ is the corresponding value computed using floating point arithmetic, then $y K^{-3^{|C|}} \leq \tilde y \leq y K^{3^{|C|}}$. If $|C| = 0$, then there are no gates in the circuit and the result trivially holds. Otherwise, consider the circuit $C'$ obtained after deleting one of the output gates. We know that the induction hypothesis holds for $C'$. 
		 Moreover, let $y$ be the exact output of the gate we just deleted, and $\tilde y$ be its approximate output computed using floating point arithmetic. Then we have:
		\begin{itemize}
			\item If the gate we deleted is an addition gate, then let $a$ and $b$ be the exact values of its input, and $\tilde a$ and $\tilde b$ the approximate values computed using floating point arithmetic. Then we know that: 
			\begin{align*}
				\tilde y = \tilde{a} \tplus \tilde{b} \leq (1+\varepsilon)(\tilde{a} + \tilde b) \leq (1+\varepsilon) K^{3^{|C| - 1}}(a + b) \leq K^{3^|C|} y \;,
			\end{align*}
			where we used the induction hypothesis on $\tilde a$ and $\tilde b$ which are positive numbers and intermediate results in the circuit $C'$. In a similar fashion, we have:
			\begin{align*}
			\tilde y = \tilde{a} \tplus \tilde{b} \geq (1-\varepsilon)(\tilde{a} + \tilde b) \geq (1-\varepsilon) K^{-3^{|C| - 1} }(a + b) \leq K^{-3^{|C|}} y \;,
			\end{align*}
			
			\item If the gate we deleted was a multiplication gate, then again let $a$ and $b$ be its exact inputs, and $\tilde a$ and $\tilde b$ the approximate inputs computed using floating point arithmetic. Then it holds that:
			\begin{align*}
			\tilde y = \tilde{a} \ttimes \tilde{b} \leq (1+\varepsilon)(\tilde{a} \times \tilde b) \leq (1+\varepsilon) K^{2\cdot 3^{|C| -1}}(a + b) \leq K^{3^{|C|}} y \;,
			\end{align*}
			and, 
			\begin{align*}
			\tilde y = \tilde{a} \ttimes \tilde{b} \geq (1-\varepsilon)(\tilde{a} \times \tilde b) \geq (1-\varepsilon) K^{-2\cdot 3^{|C| - 1} }(a + b) \geq K^{-3^{|C|}} y \;.
			\end{align*}			
			\item Finally, if the gate was a division gate, we have in a similar way:
			\begin{align*}
			\tilde y = \tilde{a} \tdiv \tilde{b} \leq (1+\varepsilon)(\tilde{a} / \tilde b) \leq (1+\varepsilon) K^{2\cdot 3^{|C| -1}}(a / b) \leq K^{3^{|C|}} y \;,
			\end{align*}
			and, 
			\begin{align*}
			\tilde y = \tilde{a} \tdiv \tilde{b} \geq (1-\varepsilon)(\tilde{a} / \tilde b) \geq (1-\varepsilon) K^{-2\cdot 3^{|C| - 1} }(a / b) \geq K^{-3^{|C|}} y \;.
			\end{align*}						
		\end{itemize}
		
		In particular, we can compute the output of the circuit up to a multiplicative factor of $K^{3^|C|} \leq 1 + 3^{|C| + 1} \varepsilon$. Let $D \leq 2^{n_{\mathrm b} \cdot n}$ be the least common multiple of all the denominators of the input vector $x$. Since the polynomial $P$ computed by $C$ has degree $d$, it follows that $D^d P(x)$ is an integer. Hence, we only need to choose $\varepsilon$ such that the error term satisfies $y D^d3^{|C| + 1} \varepsilon < \frac 1 2$. Indeed, if this holds, then if we denote by $\tilde{y}$ the output of the circuit computed using floating point arithmetic, and $y$ is the exact output, then we have:
		\[ D^d y - \frac 1 2 < D^d(1 - 3^{|C| + 1} \varepsilon) y \leq D^d \tilde{y} \leq D^d(1 + 3^{|C| + 1} \varepsilon) y < D^d y + \frac 1 2 \;. \]
		In particular, since $D^d y$ is an integer, and the fact that the interval above contains only a single integer, this value can be retrieved by rounding $D^d \tilde{y}$ to the closest integer.

		 Moreover, the inequality $yD^d3^{|C| + 1} \varepsilon < \frac 1 2$ holds as soon as $\frac 1 \varepsilon > 2 y D^d 3^{|C| +1}$. Since we have $\varepsilon = 2^{-b_{\mathrm m}}$, it is sufficient to take $b_{\mathrm m} \geq \log(3)\cdot(|C| + 1) + \log(yD^d)+ 1$ for this inequality to hold. Since $D^d y$ is a positive integer and is at most $2^{n_{\mathrm b} \cdot n \cdot d + n_{\mathrm b}}$, it follows that taking $b_{\mathrm m}$ greater than $\log(3)\cdot(|C| + 1) + n_{\mathrm b} \cdot n \cdot d + 1 $ is sufficient. This completes the proof and shows that we only need a polynomial number of bits to achieve the required precision for the exact result to be recovered at the end.
	\end{proof}

	\section{Proof of the Theorem}
	\label{sec:strongOrd}
	
	We now have all the tools we need to prove the Theorem~\ref{thm:strongOrd}. The proof first proceeds by showing how to construct a polynomial-size positive circuit computing the independence polynomial $P_G$. This circuit will be built using the strong ordering of the graph. Before we can construct the circuit for the whole graph, we start by considering the case where we restrict ourselves to a smaller class, which will then be used as a subroutine for the main construction.
	\begin{lemma}
		\label{lem:cographs}
		If $G$ is a cograph, then we can build in polynomial time a positive circuit of size $O(n)$ computing $P_G$.
	\end{lemma}
	\begin{proof}
		We will show by induction on the size of the graph $G$ that there is a positive circuit computing $P_G - 1$, which is the polynomial corresponding to the weighted number of non-empty independent sets. The induction is done using the recursive definition of cographs in Definition~\ref{def:cographs}. If $G$ contains a unique vertex $v_1$, then $P_G - 1 = x_1$, and trivially has a constant-size positive circuit.
		
		If $G$ is a disjoint union of two cographs $G_1$ and $G_2$, then any independent set of $G$ is the union of an independent set of $G_1$ and an independent set of $G_2$. From this it follows that $P_G = P_{G_1} P_{G_2}$, and consequently we have: $P_G - 1 = (P_{G_1} - 1)(P_{G_2} - 1) + (P_{G_1} - 1) + (P_{G_2} -1)$ and we can easily construct a circuit for $P_G-1$ given circuits for both $P_{G_1} - 1$ and $P_{G_2} - 1$.
		
		Finally, if $G$ is a complete join of two cographs $G_1$ and $G_2$, then any non-empty independent set of~$G$ is either a non-empty independent set of $G_1$ or a non-empty independent set of $G_2$, but not both. This implies that $P_G - 1 = (P_{G_1} - 1) + (P_{G_2} -1)$. This implies again that the circuit for $P_G-1$ can be computed from the circuits for both $P_{G_1} - 1$ and $P_{G_2} - 1$.
		
		This proves that the polynomial $P_G-1$, and hence $P_G$ has a positive circuit. This circuit has size $O(n)$ since at each step of the induction only a constant number of gates are added, and is positive since we use only addition and multiplication gates.
	\end{proof}
	
	The main element of the proof will be the following lemma which consists in showing that, up to a modification of the weight function, we can safely remove the first vertex in a strong ordering of $G$. More precisely, we want to show that if $v$ is the first vertex in the strong ordering of $G$, then there is a weight function $w'$ of $G \setminus v$ such that $\nbWIS(G, w) = K \cdot \nbWIS(G \setminus v, w')$, where K and the coefficients of $w'$ can be computed by positive circuits. If this holds, then constructing the positive circuit for the whole graph is simply a matter of iterating this procedure and removing the vertices one by one.
	
	Let us first describe in more details how this weight function $w'$ can be constructed. Let $G$ be a strongly orderable graph, with $v$ the first vertex in a strong ordering of $G$, and $w$ a weight function on $G$. By Lemma~\ref{lem:strongOrd} we know that $H = G[N(v), N^2(v)]$ induces a chain graph. We denote by $v_1, \ldots, v_k$ the vertices in $N(v)$ ordered according to the inclusion of their neighbourhoods in $H$. In other words, $N_H(v_{i}) \subseteq N_H(v_{i+1})$. Let us denote by $G_i$ the graph induced by $\{v_1, \ldots v_{i-1} \} \setminus N(v_i)$. We will take the convention that $G_0$ is the empty graph, and $G_k = G[N(v)]$. We consider the weight function $w'$ on $G \sm v$ such that $w'$ and $w$ agree on $G \sm N[v]$, and for all the vertices in $N(v)$: 
	\begin{align}
	w'(v_i) = w(v_i)\cdot (1 + w(v)) \cdot \frac {\nbWIS(G_i, w|_{G_i})}{\nbWIS(G_i, w'|_{G_i})} \;. \label{eq:wdef}
	\end{align}
	
	Note that in this definition, $w'(v_i)$ depends only on $w'(v_j)$ for $j < i$, and consequently this defines $w'$ uniquely. Moreover, since $G_i$ is a cograph, then by Lemma~\ref{lem:cographs} it follows immediately that the coefficients of $w'$ can be computed using only positive circuits. Moreover, the following result holds:
	\begin{lemma}
		\label{lem:equivWeights}
		We have $\nbWIS(G, w) = (1 + w(v)) \cdot \nbWIS(G \sm v, w')$.
	\end{lemma}
	\begin{proof}
		The proof proceeds by partitioning the set of independent sets of $G$, $\indepSet(G)$, depending on how the independent sets intersect the vertex set of $G \sm N[v]$. More precisely, let us denote by $\cI_j$ the set of independent sets $I \in \indepSet(G \sm N[v])$ such that $j$ is the largest index such that $v_j$ is not a neighbour of any vertex in $I$, with the convention that $\cI_0$ are the independent sets adjacent to all the vertices of $N(v)$. Note that the $(\cI_j)_{0 \leq j \leq k}$ is a partition of of $\indepSet(G \sm N[v])$.
		
		We can remark that if an independent set $I$ of $G \sm N[v]$ is in $\cI_j$, then the vertices of $N[v]$ it is not adjacent to are exactly $v_1, \ldots, v_j$. Indeed, since $G[N(v), N^2(v)]$ induces a chain graph, then any vertex $w \in G \sm N[v]$ that is adjacent to $v_j$ is also adjacent to $v_i$ for all $i \geq j$. Hence, it follows from this observation that we can write 
\[ \indepSet(G) = \bigcup_{j = 0}^k \{S \cup I \mid S \in \indepSet(G[\{v, v_1, \dots, v_j \}]) \text{~and~} I \in \cI_j \} \;. \]
In other words, this means that the independent sets of $G$ are obtained by extending for all $j \geq 0$ the elements of $\cI_j$ with an independent set of $G[\{v, v_1, \ldots, v_j\}]$. And conversely, any set obtained with this method is an independent set of $G$. Hence we have
		\begin{align*}
		\nbWIS(G, w) &= \sum_{j = 0}^k \alpha_j \sum_{I \in \cI_j} \prod_{u \in I} w(u) \;,
		\end{align*}
		where $\alpha_j = \nbWIS(G[\{v, v_1, \ldots, v_j\} ])$. Again, an independent set $I \in \indepSet(G[\{v_, v_1, \ldots , v_j\} ])$ can be one of the following:
		\begin{itemize}
			\item $I$ is the empty set, 
			\item $I$ is a singleton containing only $v$, 
			\item or $I$ is an independent set containing $v_i$ for some $i \leq j$ but no $v_{i'}$ for any $i' > i$.
		\end{itemize}
		This case distinction implies that the coefficient $\alpha_j$ can be rewritten as:
		\[ \alpha_j = 1 + w(v) + \sum_{i = 1}^j w(v_i)\cdot \nbWIS(G_i, w) \;. \]
		In this sum, the $1$ corresponds to the empty set, and the term $w(v)$ corresponds to the independent set containing only $v$.
		
		Using exactly the same argument as above for the graph $G -v$ using the weight function $w'$, we can write in a similar fashion:
		\[ \nbWIS(G \sm v, w') = \sum_{j = 0}^k \alpha'_j \sum_{I \in I_j} \prod_{u \in I} w(u) \;, \]
		where $\alpha'_j = \nbWIS(G[\{v_1, \ldots v_j \}])$. Again, these coefficients satisfy the equality:
		\[ \alpha'_j =  1  + \sum_{i = 1}^j w'(v_i) \nbWIS(G_i, w') \;. \]
		
		Hence, to prove the result, it is enough to show that for all $j$, $\alpha'_j = (1 +w(v)) \alpha_j$, which is equivalent to proving that for all $i$, we have $w'(v_i) \cdot \nbWIS(G_i, w') = (1 + w(v)) \cdot w(v_i) \cdot \nbWIS(G_i, w)$, which corresponds exactly to the definition of $w'$ from Equation~(\ref{eq:wdef}).
	\end{proof}
	
	\begin{corollary}
		\label{cor:buildCircuit}
		There is an algorithm that, given a strongly orderable graph $G$, outputs a (polynomial size) positive circuit for $P_G$ in polynomial time.
	\end{corollary}
	\begin{proof}
		The proof follows from applying the result of Lemma~\ref{lem:computeCircuit} repeatedly. More precisely, we show the result by induction on the number of vertices in $G$. If $G$ is empty, $P_G = 1$ and the result trivially holds. Otherwise, let $v$ be the first vertex in a strong ordering of $G$. Using the induction hypothesis, we can build a positive circuit $C'$ which computes $P_{G \sm v}$. Then, we know that for any weight function $w$, we can define a weight function $w'$ as in Equation~(\ref{eq:wdef}), and it follows from Lemma~\ref{lem:equivWeights} that $\nbWIS(G,w) = (1 + w(v)) \nbWIS(G \sm v, w')$. We know by Lemma~\ref{lem:strongOrd} that $G[N(v)]$ is a cograph. Hence, by definition of $w'$, and using Lemma~\ref{lem:cographs}, it follows that the coefficients of $w'$ can be computed with a positive circuit from the coefficients of $w$. Plugging this into the inputs of the circuit $C'$, and dividing the result by $1 + w(v)$ creates a positive circuit which computes $P_G$.
	\end{proof}

	The proof of the main theorem then follows immediately by combining the different lemmas together.
	\begin{proof}[Proof of Theorem~\ref{thm:strongOrd}.]
		By Corollary~\ref{cor:buildCircuit}, we can build in polynomial time a positive circuit $C$ which computes $P_G$. Since $P_G$ has degree at most $n$, it follows from Lemma~\ref{lem:computeCircuit} that we can evaluate exactly and in polynomial time the output of the circuit $C$ on any positive weight function $w$, and the result follows.
	\end{proof}

	\section{Graphs of bounded clique-width}
	\label{sec:boundedCW}
	
	We now prove that counting the weighted number of independent sets can be done in polynomial time for graphs of bounded clique-width. Let us start by giving a formal definition of the clique-width of a graph before proving this result.
	
Let $\Lambda$ be a finite set of labels and let $U$ be a (countable) set of
vertices. Recursively we define \textbf{\LE s} and the labelled graphs they
describe. Here a \textbf{labelled graph} is a triple $(V,E,\lambda)$ where
$(V,E)$ is a graph and $\lambda: V \to \Lambda$ assigns labels to the
vertices. \LE s are built using the four following operations:
	
\begin{description}
\item[Create a vertex.] For every label $i \in \Lambda$ and every vertex $v
  \in U$ the string $\bullet_i(v)$ is a \LE. It describes the labelled graph
  $(\{v\}, \varnothing, \lambda)$ with $\lambda(v)=i$.
\item[Disjoint union.] Let $\sigma$ and $\tau$ are \LE s describing
  $(V,E,\lambda)$ and $(W,F,\mu)$. If $V \cap W = \es$ then $\sigma \oplus \tau$
  is a \LE. It describes the disjoint union of the two graphs described by 
  $\sigma$ and $\tau$, in other words the graph $(V \cup W, E \cup F, \gamma)$ where:
  \[ \gamma(v) = \begin{cases}
                   \lambda(v) & \text{if~} v \in V, \\
		   \mu(v)     & \text{if~} v \in W. 
		  \end{cases} \]
\item[Add edges.] For different labels $i,j \in \Lambda$ and a \LE\ $\sigma$
  the string $\eta_{i \udtimes j}(\sigma)$ is a \LE. It describes the
  graph where all the edges between colours $i$ and $j$ were added. Formally,
  if~$\sigma$ corresponds to the graph $(V, E, \lambda)$ then 
  $\eta_{i \udtimes j}(\sigma)$ describes $(V, E \cup F, \lambda)$ where 
  $F = \{vw \mid \lambda(v)=i \text{~and~} \lambda(w)=j\}$.
\item[Relabel.] For different labels $i,j \in \Lambda$ and a \LE\ $\sigma$ the
  string $\rho_{i \to j}(\sigma)$ is a \LE\ describing the graph were vertices
  labelled $i$ were relabelled to $j$. Formally, if $\sigma$ describes $(V, E,
  \lambda)$ then $\rho_{i \to j}(\sigma)$ describes $(V, E, \mu)$ where:
  \[ \mu(v) = \begin{cases}
		j          & \text{if~} \lambda(v)  =  i, \\ 
		\lambda(v) & \text{if~} \lambda(v) \ne i.
	       \end{cases} \]
\end{description}
	
The \textbf{clique-width} of a graph $G=(V,E)$, denoted by $\cwd(G)$, is the
minimum size of a set $\Lambda$ such that there is a labelling $\lambda: V \to
\Lambda$ and a \LE\ describing $(V,E,\lambda)$.
	
If we additionally allow the empty string to be a \LE\ then $(\es,\es)$ is the only graph of clique-width zero. A graph $G=(V,E)$ has clique-width at most one if and only if $E=\es$. The graph $G$ has clique-width at most two if and only if $G$ is a cograph.
There is a polytime algorithm recognising graphs of clique-width at most three \cite{CHLRR12}. In general it is \NP-complete to decide whether for given $G$ and $k$, $\cwd(G) \le k$ holds \cite{FRRS06}. In contrast to tree-width, the parameterised problem for clique-width is not known to be fixed parameter tractable.
	
The result we want to show is the following
\begin{theorem}
  There is an algorithm which, given a graph $G$, a \LE\ of $G$ with
  $|\Lambda| = \ell$ and a weight function $w$, computes $\nbWIS(G, w)$ in
  time $O(2^\ell \ell^2 \poly(n))$.
\end{theorem}
The exponent in the polynomial in the result above is independent of the bound $\ell$ on the clique-width of the graph.
	This result implies Theorem~\ref{thm:boundedCW}. Indeed, if the \LE\ is not given as input to the algorithm, then using the result from~\cite{OS06}, a \LE\ with $|\Lambda| \leq (2^{3\cwd(G) + 1} +1)$ can be computed in polynomial time (if $\cwd(G)$ is constant). This implies that the algorithm above is still polynomial on graphs of bounded clique-width, even when the corresponding expression is not given to the algorithm, however the dependency in the clique-width then becomes a double exponential.
	
	\begin{proof}
	Let us start with a few notations. In all the rest of this proof, $\Lambda$ will be a finite set of labels, and given a \LE\ $\sigma$, we will denote by $G_{\sigma}$ the labelled graph described by this expression. Moreover, if $H$ a labelled graph with a labelling $\lambda$, and $\Gamma \subseteq \Lambda$ is a subset of labels, then $H\an{\Gamma}$ denotes the subgraph induced by the vertices with a label in $\Gamma$, in other words: $H\an{\Gamma} = H[\{v \mid \lambda(v) \in \Gamma\}]$.
	
	Let $G$ be the input graph with the weight function $w$, and let $\mu$ be the \LE\ describing $G$. The algorithm will proceed using dynamic programming by computing, for some \LE\ $\sigma$ and for every different subsets of labels $\Gamma \subseteq \Lambda$ the quantity $\nbWIS(G_{\sigma} \an{\Gamma}, w)$ that will be denoted by  $c(\sigma, \Gamma)$ to simplify notations. These quantities will be computed by induction using the recursive definition of the \LE s. 
	 
	The subexpressions of $\mu$ form a decomposition tree. Its leaves are the $\bullet$-nodes, exactly one for each vertex of $G$. Moreover the tree contains $n-1$ inner nodes with two children, these are the $\oplus$-nodes. All other nodes have exactly one child and are $\eta$- or $\rho$-nodes. There is always a decomposition tree such that each path between two $\oplus$-nodes (or an $\oplus$-node and the root) has $O(|\Lambda|^2)$ nodes of $\eta$- or $\rho$-type. In short, we can assume that the size of the decomposition tree is $O(|\Lambda|^2n)$.
	
	Let $\sigma$ be a subexpression of $\mu$, and $\Gamma \subseteq \Lambda$ be a subset of labels. Let us show how $c(\sigma, \Gamma)$ can be computed using the recursive structure of \LE s:
	
	\begin{description}
		\item[Create a vertex.] Assume first that $\sigma = \bullet_i(v)$ for some label $i \in \Lambda$ and a vertex   
		$v \in U$. Then we have:
		\begin{align*}
		c(\sigma, \Gamma) = \begin{cases}
			1 + w(v) & \text{if~} i    \in \Gamma, \\
			1        & \text{if~} i \notin \Gamma.
		\end{cases}
		\end{align*}
		Indeed, if $i \in \Gamma$ then $G$ contains a single vertex $v$, and there are two possible independent sets: the empty set and the set containing $v$, and if $i \not \in \Gamma$, then $G$ is the empty graph and the empty set is the only independent set of $G$.
		
		\item[Disjoint union.] Assume that $\sigma = \tau_1 \oplus \tau_2$, then we have:
		\[ c(\sigma, \Gamma)  = c(\tau_1, \Gamma)  \cdot c(\tau_2,\Gamma) \;. \]
		Indeed, since $G_{\sigma}$ is the disjoint union of $G_{\tau_1}$ and $G_{\tau_2}$, any independent set of $G_\sigma$ is the union of an independent set of $G_{\tau_1}$ and an independent set of $G_{\tau_2}$. And reciprocally, every set of this form is an independent set of $G_{\sigma}$. This holds even when the graph is restricted to the vertices with a label in $\Gamma$.

		\item[Add edges.] Assume now that $\sigma = \eta_{i \udtimes j}(\tau)$. Each independent set of $G_\sigma\an{\Gamma}$ is independent
		in $G_\tau\an{\Gamma\setminus\{i\}}$ or $G_\tau\an{\Gamma\setminus\{j\}}$ and the independent sets
		of $G_\tau\an{\Gamma\setminus\{i,j\}}$ are independent both in $G_\tau\an{\Gamma\setminus\{i\}}$ 
		and in $G_\tau\an{\Gamma\setminus\{j\}}$. Therefore we have
		\[ c(\sigma,\Gamma) = c(\tau,\Gamma \setminus \{i\}) + c(\tau,\Gamma \setminus \{j\}) - c(\tau,\Gamma \setminus \{i, j\}) \;. \]
		
		\item[Relabel.] Finally, assume that $\sigma = \rho_{i \to j}(\tau)$. Then we have:
	     \[ c(\sigma, \Gamma) = 
             \begin{cases}
		c(\tau, \Gamma \sm  \{i\}) & \text{if~} j \notin \Gamma, \\
		c(\tau, \Gamma \cup \{i\}) & \text{if~} j    \in \Gamma.
             \end{cases} \]
		This follows from the fact that relabelling the vertices only change the subset of labels $\Gamma$ under consideration. 
	\end{description}
	
	A dynamic programming algorithm evaluating this recurrence will perform at most $O(2^{\ell}\ell^2n)$ steps where $\ell = |\Lambda|$. Each step consists only in performing some arithmetic operations on numbers represented by a polynomial number of bits, hence the whole algorithm runs in polynomial time, which completes the proof of the theorem.
	\end{proof}

	\section{Conclusion}
	We have shown that the number of independent sets of a graph can be computed in polynomial time for strongly orderable graphs and graphs of bounded clique-width. One natural question is whether these results can be generalized further. A natural class of graph to consider is the class of weakly chordal graphs for which the complexity is still open. One other remark is that our algorithm for bounded clique-width graphs is doubly exponential in the clique-width of the graph when a lambda expression is not given as input to the algorithm. This is due to the fact that there is currently no FPT algorithm to compute the clique-width of a graph. One way to get around this problem could be to try to adapt the algorithm for the rank-width, a parameter closely related to clique-width which can be computed in FPT time.

\section*{Acknowledgement}

The authors are grateful to Martin Dyer for stimulating discussions.

	\bibliographystyle{alpha}
	\bibliography{bib}

\end{document}